\documentclass{PoS}
\usepackage{amsmath}
\usepackage{subcaption}
\usepackage{verbatim}

\newcommand{\cev}[1]{\reflectbox{\ensuremath{\vec{\reflectbox{\ensuremath{#1}}}}}}

\title{Finite volume renormalization scheme for fermionic operators}

\ShortTitle{Finite volume renormalization scheme for fermionic operators}

\author{Christopher Monahan\\
        The College of William \& Mary\\
        E-mail: \email{cjmonahan@wm.edu}}

\author{\speaker{Kostas Orginos}\\
       The College of William \& Mary,\\Thomas Jefferson National Accelerator Facility\\
       E-mail: \email{kostas@jlab.org}}

\abstract{We propose a new finite volume
renormalization scheme. Our scheme is based on the
Gradient Flow applied to both fermion and gauge fields and, much like the 
Schr\"odinger
functional method, allows for a nonperturbative determination of the
scale dependence  of operators using a step-scaling approach. We give some
preliminary results for the pseudo-scalar density in the quenched 
approximation.
}

\FullConference{31st International Symposium on Lattice Field Theory LATTICE 
2013\\
		 July 29, August 3, 2013\\
		 Mainz, Germany}

\begin{document}

\section{Introduction}
 
Renormalization is a major problem for calculations of moments of generalized 
parton distribution functions (GPDs) in lattice QCD. 
The lattice breaks rotational symmetry, causing operators of different mass 
dimension to mix. This mixing complicates
the renormalization of high moments of GPDs and therefore only calculations 
of a few low moments exist in the literature.

Smearing has long been used as a tool to aid renormalization 
on the lattice~\cite{Bernard:1999kc,Narayanan:2006rf}. Recent developments of 
Gradient Flow (GF) methods~\cite{Luscher:2010iy,Luscher:2011bx,Luscher:2013cpa} 
in both the gauge and fermion sectors offer an opportunity to improve current 
renormalization methods for the twist-2 operators relevant to moments of 
GPDs. Using smearing to alleviate the mixing of lattice twist-2 operators with 
lower dimensional operators was recently discussed 
in~\cite{Davoudi:2012ya}.

In this work we explore the possibility of using the Gradient Flow to define a 
new renormalization scheme.
 This nonperturbative scheme potentially provides a simple way to calculate 
the mixing coefficients required to subtract power divergent lower 
dimensional operators from lattice twist-2 matrix elements.  
We propose a method that extends the approach taken in~\cite{Fodor:2012td}, 
which uses the Gradient Flow to define a renormalized 
QCD coupling constant in finite volume, to compute
  the scale dependence of twist-2 operators as well as the mixing coefficients. 
Here we outline our proposal and, as a first step
  towards understanding the effectiveness of our approach, we compute the 
step-scaling function for the pseudo-scalar density.

\section{The Gradient Flow}
L\"uscher introduced the Gradient Flow method for both 
fermions and gauge fields, which corresponds to 
smearing all degrees of freedom in a new, continuous flow~time direction 
$s$~\cite{Luscher:2013cpa}. 
The flow equations can be written as
\begin{equation}
\partial_s V_\mu(x,s) = -g_0^2\partial_{V_\mu(x,s)} S_w V_\mu(x,s),\quad 
\partial_s \psi(x,s) =  \vec{\Delta} \psi(x,s),\quad \mathrm{and}\quad
\partial_s \overline{ \psi}(x,s) =  \overline{\psi}(x,s) \cev{\Delta},
\end{equation}
where $S_w$ is the Wilson gauge action and the derivative 
$\partial_{V_\mu(x,s)}$, taken with respect to the smeared gauge links 
$V_\mu(x,s)$, is defined as in~\cite{Luscher:2010iy}. We denote the smeared 
fermion fields $\psi(x,s)$ and $\overline{\psi}(x,s)$, 
and define the lattice Laplacian as $\Delta =  \nabla^{\dagger}_\mu 
\nabla_\mu$, where 
$\nabla_\mu$ is the covariant backward lattice derivative. 

The smeared fields satisfy the 
boundary conditions
\begin{equation}
V_\mu(x,0)=U_\mu(x),\qquad \psi(x,0)=q(x),\quad 
\mathrm{and}\quad\overline{\psi}(x,0)=\overline{q} (x).
\end{equation}
Here the $U_\mu(x)$ are the unsmeared gauge fields and $\overline{q}(x)$ and 
$q(x)$ the unsmeared quark fields.
 
If the underlying theory, defined as a path integral over the unsmeared 
fields, is renormalized, then correlation functions of the gauge
fields require no additional renormalization, as shown 
in~\cite{Luscher:2011bx}. At finite flow time the smeared fermion fields 
require one additional wave function renormalization~\cite{Luscher:2013cpa}. 
One way to easily understand this is to notice that the effective 
smearing range, which is proportional to $\sqrt{s}$, acts as a regulator that
eliminates additional divergences at non-zero flow time.
 
\section{Finite Volume Renormalization Scheme}
We define the renormalization constant $Z_{{\cal O}_\Gamma}(\mu)$  of an 
operator ${\cal O}_{\Gamma}$ at a scale $\mu$ as\vspace*{-0.2\baselineskip}
\begin{equation}
Z_{{\cal O}_\Gamma}(g_0,\mu) = {\cal N} \frac{\sum_x \langle V_\nu(x,s) 
V_\nu(0,0)\rangle}{\sum_x \langle {\cal O}_\Gamma(x,s) {\cal 
O}_\Gamma(0,0)\rangle},
\label{eq:renormFact}
\end{equation}
where $g_0$ is the bare gauge coupling constant and
\begin{equation}
V_\nu(x,s) = \bar \psi(x,s) \gamma_\nu \psi(x,s)\, ,\qquad \mathrm{and}\qquad 
{\cal 
O}_{\Gamma}(x,s) = \bar\psi(x,s) \Gamma \psi(x,s).
\end{equation}
Here $\Gamma$ is some gamma matrix.

We choose a novel normalization factor, ${\cal 
N}$, that is given by
\begin{equation}
{\cal N} = \frac{\sum_x \langle {\cal O}_\Gamma(x,s) {\cal 
O}_\Gamma(0,0)\rangle_{{\rm cg}}}{\sum_x \langle V_\nu(x,s) 
V_\nu(0,0)\rangle_{{\rm cg}}}\,,
\label{eq:normFact}
\end{equation}
where the subscript ${\rm cg}$ signifies that we evaluate the
expectation value of the matrix element over an ensemble of  
constant gauge fields. Our choice of normalization removes the need to account 
for gauge 
zero modes, which would otherwise require a nonperturbative treatment when 
matching 
perturbatively to the $\overline{MS}$ scheme.

We calculate the correlation 
functions in 
finite volume of size $L^4$ and set $\mu=1/L$. Following~\cite{Fodor:2012td} 
we set the flow time to
\begin{equation}
s = c^2\; \frac{L^2}{8},
\end{equation}
where $c$ is a constant that defines the renormalization scheme. In this work we
investigate a range of values for the constant $c$. The optimum value is 
$c\simeq 0.3$, as discussed in \cite{Fodor:2012td}, and is determined by the 
interplay of statistical errors and systematic uncertainties from cut-off 
effects. On the one hand, large values of 
$c$ increase statistical errors in the determination of the renormalization 
factor. On the other hand, small values of $c$ 
lead to renormalized couplings that have large discretization errors and are 
only weakly dependent on the bare 
coupling (see Figure \ref{fig:gbar}), which entails a prohibitively large 
spread of bare coupling values for the step-scaling procedure.
 
We would like to highlight the four following features of our definition of 
the renormalization factor, $Z_{{\cal O}_\Gamma}(g_0,\mu)$, in Equation 
\eqref{eq:renormFact}.\vspace*{-0.2\baselineskip}
\paragraph{Mixed smeared and unsmeared fields} The denominator in 
Equation~\eqref{eq:renormFact} is a correlation function between the local 
(un-smeared) quark bilinear operator, ${\cal O}_\Gamma(0,0)$, and the smeared 
operator at some non-zero flow time, ${\cal 
O}_\Gamma(x,s)$.\vspace*{-0.2\baselineskip}

\paragraph{Wave function renormalization cancellation} This 
correlation function only requires a renormalization factor for the local 
quark bilinear. In principle the smeared quark bilinear requires an additional 
wave function renormalization, as discussed 
in~\cite{Luscher:2013cpa}. We have, however, chosen the numerator of  
Equation \eqref{eq:renormFact} so that the wave 
function renormalization parameters at both zero and non-zero flow time 
cancel.\vspace*{-0.2\baselineskip}

\paragraph{Constant field normalization} We adopt the renormalization 
condition of Equation \eqref{eq:normFact} so that 
the perturbative expansion of the renormalization constant receives no 
contributions at tree level from insertions of the gauge zero modes. This 
simplifies the perturbative expansion in finite volume at one loop level, at 
the cost of having to evaluate the contributions from zero 
modes nonperturbatively. The authors of \cite{Fodor:2012td} treat these zero 
mode contributions 
analytically. We evaluate Equation \eqref{eq:normFact} numerically. We 
work in a finite volume with anti-periodic boundary conditions for fermions, so
there are no fermionic zero modes and correlation functions can be evaluated 
with massless fermions. For this exploratory work we use Wilson 
fermions and fix the bare quark mass to the critical 
mass.\vspace*{-0.2\baselineskip}

\paragraph{Local vector current} We use the local vector current in 
Equations~\eqref{eq:renormFact} and \eqref{eq:normFact}. We intend to use the
conserved vector current in the future, because this simplifies the 
renormalization condition. This work is a proof of principle calculation, 
so it is 
sufficient to study our renormalization scheme using the 
simpler local vector current.

\section{Coupling constant and operator scale dependence}

Following the approach taken in \cite{Fodor:2012td}, we examine the 
discrete $\beta$-function, defined as
\begin{equation*}
{\cal B}(\mu) = \frac{\overline{g}^2(\mu/2) - \overline{g}^2(\mu)}{2\log 2}
\end{equation*}
for a step size of two. Here $\overline{g}^2(\mu)$ is the renormalized 
coupling defined in \cite{Fodor:2012td}. Our numerical results can be 
compared to the universal $\beta$-function at one 
loop in perturbation theory:
${\cal B}(\mu) = 11\overline{g}^4(\mu)/(16\pi^2)$.
The relation 
between $\overline{g}^2(\mu)$ and, for example, the $\overline{MS}$ coupling,  
contains odd powers of $\overline{g}(\mu)$. Therefore, as noted  in 
\cite{Fodor:2012td}, only the one loop coefficient is scheme independent. 
Higher order corrections to the perturbative 
$\beta$-function have not been calculated yet, 
so direct comparison with analytic results is not possible.

In common with other finite volume renormalization schemes, such as the 
Schr\"odinger functional method, we study the scale dependence of the 
renormalization parameter $Z_{{\cal O}_\Gamma}(\mu)$ via the continuum
step-scaling function $\sigma$. 
We use a step factor of two for an 
operator ${\cal O}$ and first define the discrete step-scaling function, 
$\Sigma$, as:
\begin{equation}\label{eq:sigdis}
\Sigma(\overline{g}^2(\mu),a/L) = \frac{Z_{\cal O}(g_0,\mu/2,a/L)}{Z_{\cal 
O}(g_0,\mu,a/L)}.
\end{equation}
Then the continuum step-scaling 
function $\sigma$ is
\begin{equation}\label{eq:sigcon}
\sigma(\overline{g}^2(\mu)) = \lim_{a/L\rightarrow 
0}\Sigma(\overline{g}^2(\mu),a/L).
\end{equation}

\section{Numerical tests}

\begin{figure}[t]
\begin{subfigure}{.5\textwidth}
  \centering
\includegraphics[width=\textwidth,height=0.9\textwidth]{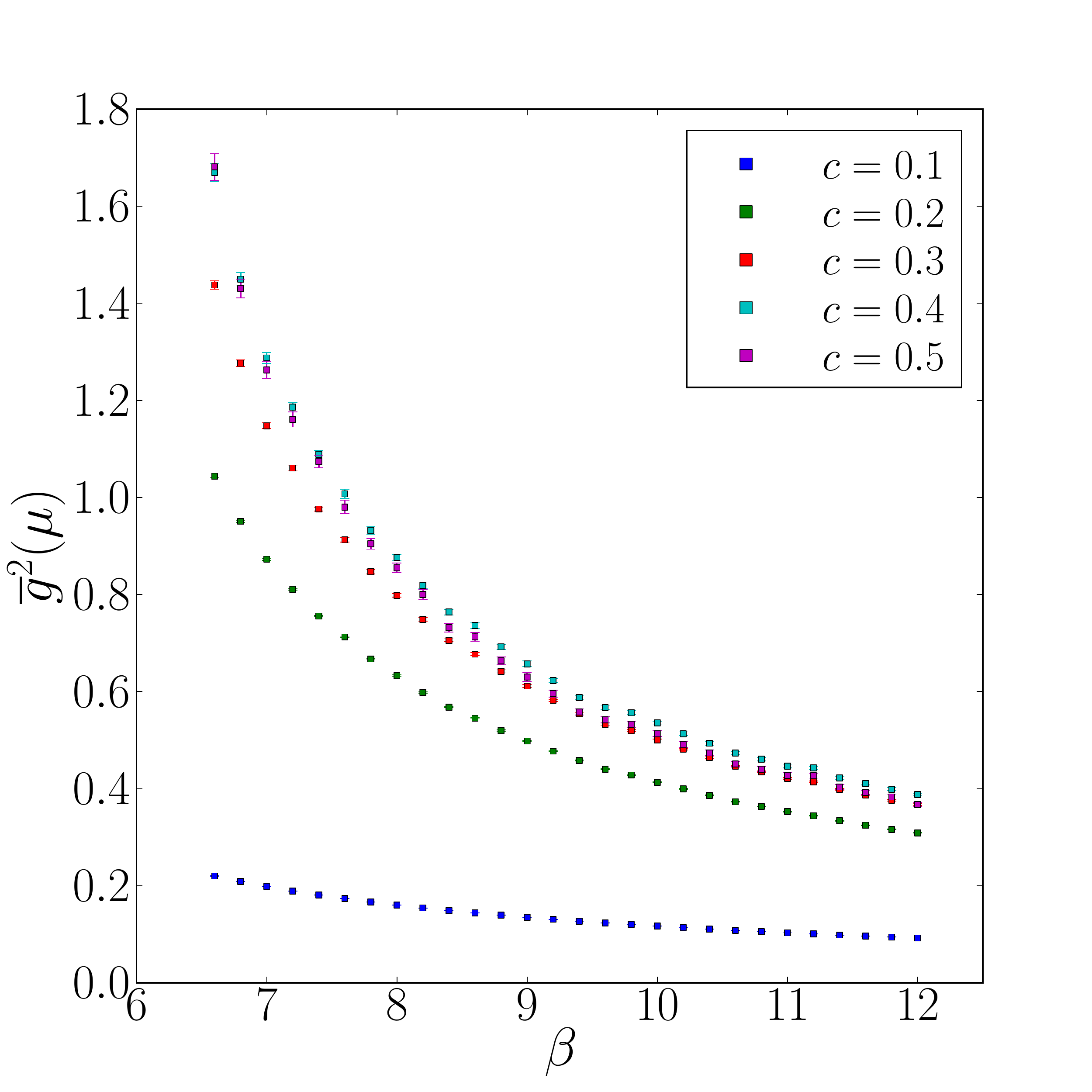}
\end{subfigure}
\begin{subfigure}{.5\textwidth}
  \centering  
\includegraphics[width=\textwidth,height=0.9\textwidth]{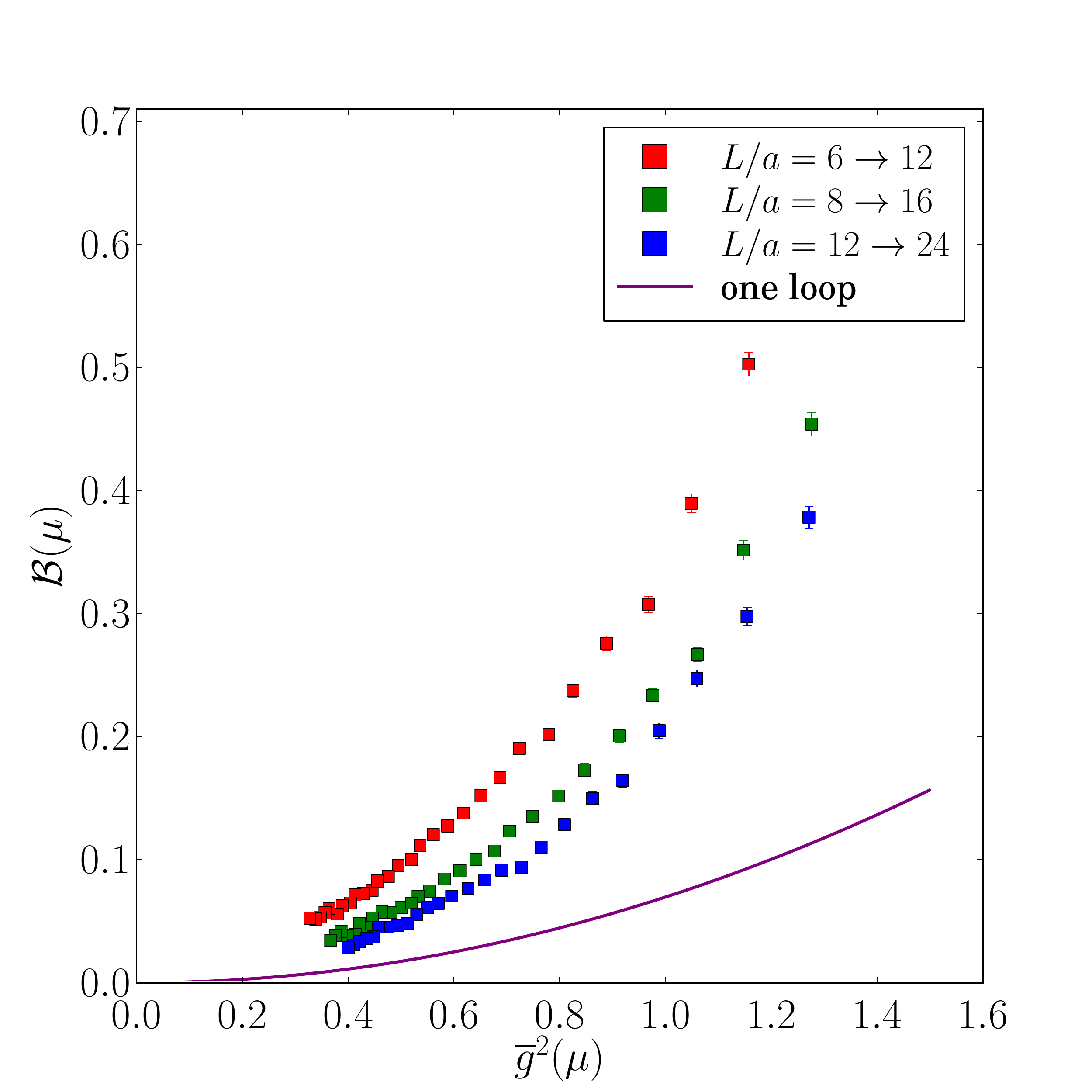}
\end{subfigure}
   \caption{ (Left) Dependence of the renormalized gauge coupling on the choice 
of the $c$ parameter at fixed lattice extent $L/a=8$. (Right) The discrete 
$\beta$-function, ${\cal B}(\mu)$, as a function of the renormalized 
coupling. Here the step size is two and we plot three different pairs of 
lattice extents with constant $c=0.3$. We include the one loop
prediction for comparison (solid purple line). }
  \label{fig:gbar}
\end{figure}

We calculated the discrete $\beta$-function and the step-scaling function for 
the pseudoscalar density using the Wilson gauge and fermion actions in the 
quenched approximation.
We generated ensembles for eight lattice sizes, from $L/a = 6$ to $L/a 
= 24$, with a range of bare couplings, from $\beta = 6/g_0^2 = 6.6$ to $\beta
= 12$. Our analysis uses ensembles with 100 configurations for each lattice volume, generated with 
500 updates of one heat bath step followed by three steps of over-relaxation.

We determined the renormalized coupling in the gradient flow finite 
volume scheme of \cite{Fodor:2012td}. We present our results as a 
function of the bare coupling, $\beta$, in Figure \ref{fig:gbar} for different 
choices of the 
constant $c$ (left-hand plot). In the right-hand plot of Figure \ref{fig:gbar} 
we plot the discrete $\beta$-function for this coupling constant, with the one 
loop prediction shown for comparison. Deviations from this prediction at small 
coupling represent discretization effects.

We determine the continuum step-scaling 
function of Equation \eqref{eq:sigcon} as follows:\vspace*{-8pt}
\begin{enumerate}
\setlength{\itemsep}{-2pt}
\item  Calculate $\overline{g}^2(\mu)$ and the pseudoscalar 
renormalization parameter, $Z_p(g_0,\mu,a/L)$, at a fixed choice of bare 
coupling, 
say $\beta = 6.6$, for a specific lattice, say, $L/a = 6$. This sets the 
physical scale, $\mu=1/L$.
\item Keep the bare coupling (and therefore the lattice spacing) fixed,
double the number of lattice points, \emph{e.g.}~set $L^\prime/a = 
2L/a=12$, and calculate $Z_p(g_0,\mu/2,a/L)$.
\item Determine the discrete step-scaling function using Equation 
\eqref{eq:sigdis}, \emph{i.e.}~calculate 
the ratio $Z_p(g_0,\mu/2,a/L)/Z_p(g_0,\mu,a/L)$.
\item Now fix the renormalized coupling. For the example of $L/a = 6$ with 
$\beta = 6.6$, this is $\overline{g}^2(\mu) = 1.158(7)$ . Choose a new lattice 
extent, say $L/a^\prime = 8$ and tune the bare coupling such that the 
renormalized coupling is constant, $\overline{g}^2(\mu) =1.158(7)$. This 
adjusts the lattice spacing on the new lattice so that the physical extent 
remains constant.
\item Repeat steps one to four.
\item Extract the continuum step-scaling function, Equation \eqref{eq:sigcon}, 
from a linear fit to $a/L$ using the results of steps one to 
give.
\item Repeat steps one to six with a different initial choice of bare 
coupling, say $\beta = 7.0$, which sets another initial physical scale 
$\mu^\prime$. Repeated application of this algorithm gives the continuum 
step-scaling function at a range of physical scales.
\end{enumerate}

\begin{figure}[t]
\begin{subfigure}{.5\textwidth}
  \centering
\includegraphics[width=\textwidth,height=0.9\textwidth]{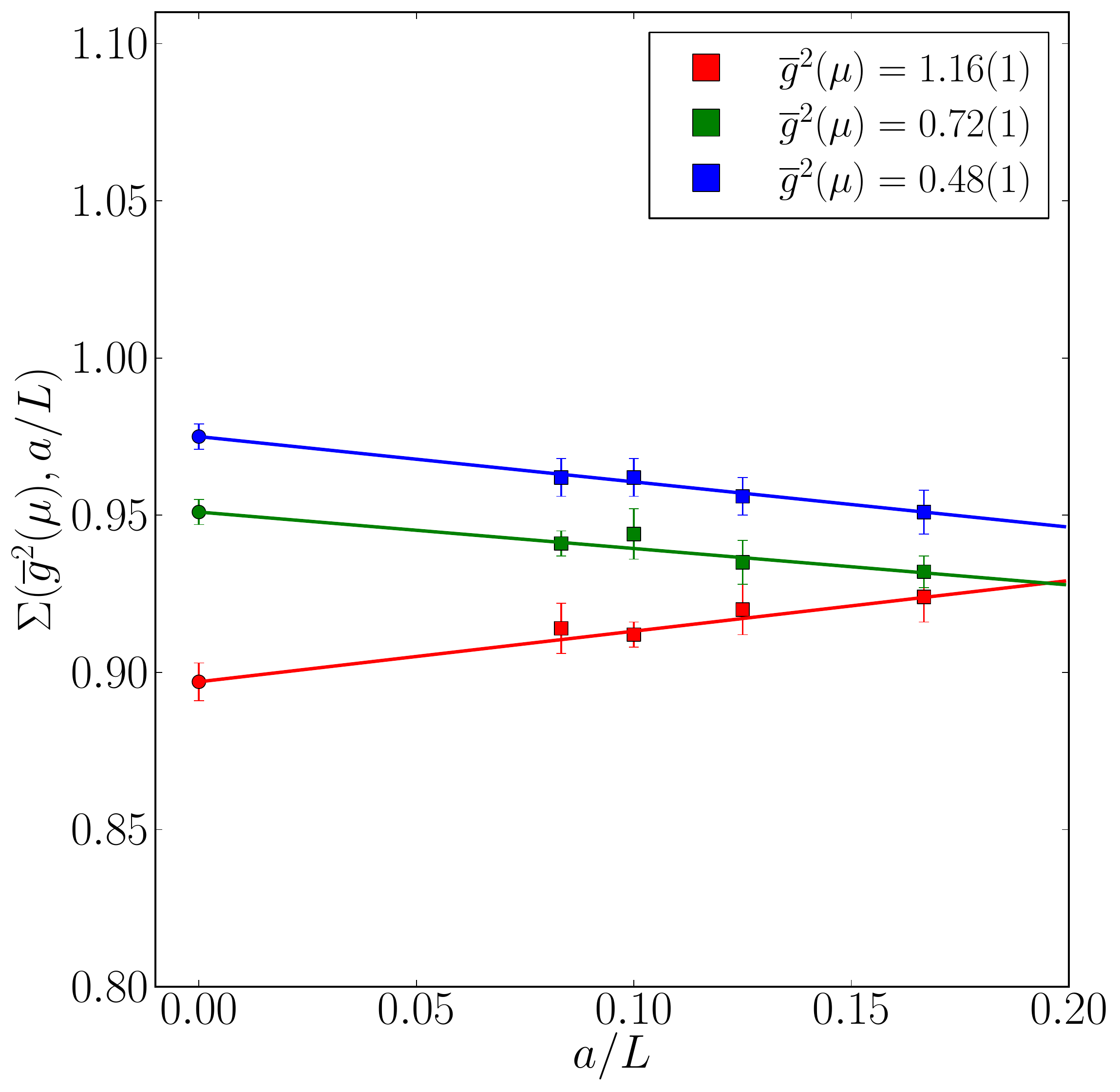}
\end{subfigure}
\begin{subfigure}{.5\textwidth}
  \centering  
\includegraphics[width=\textwidth,height=0.9\textwidth]{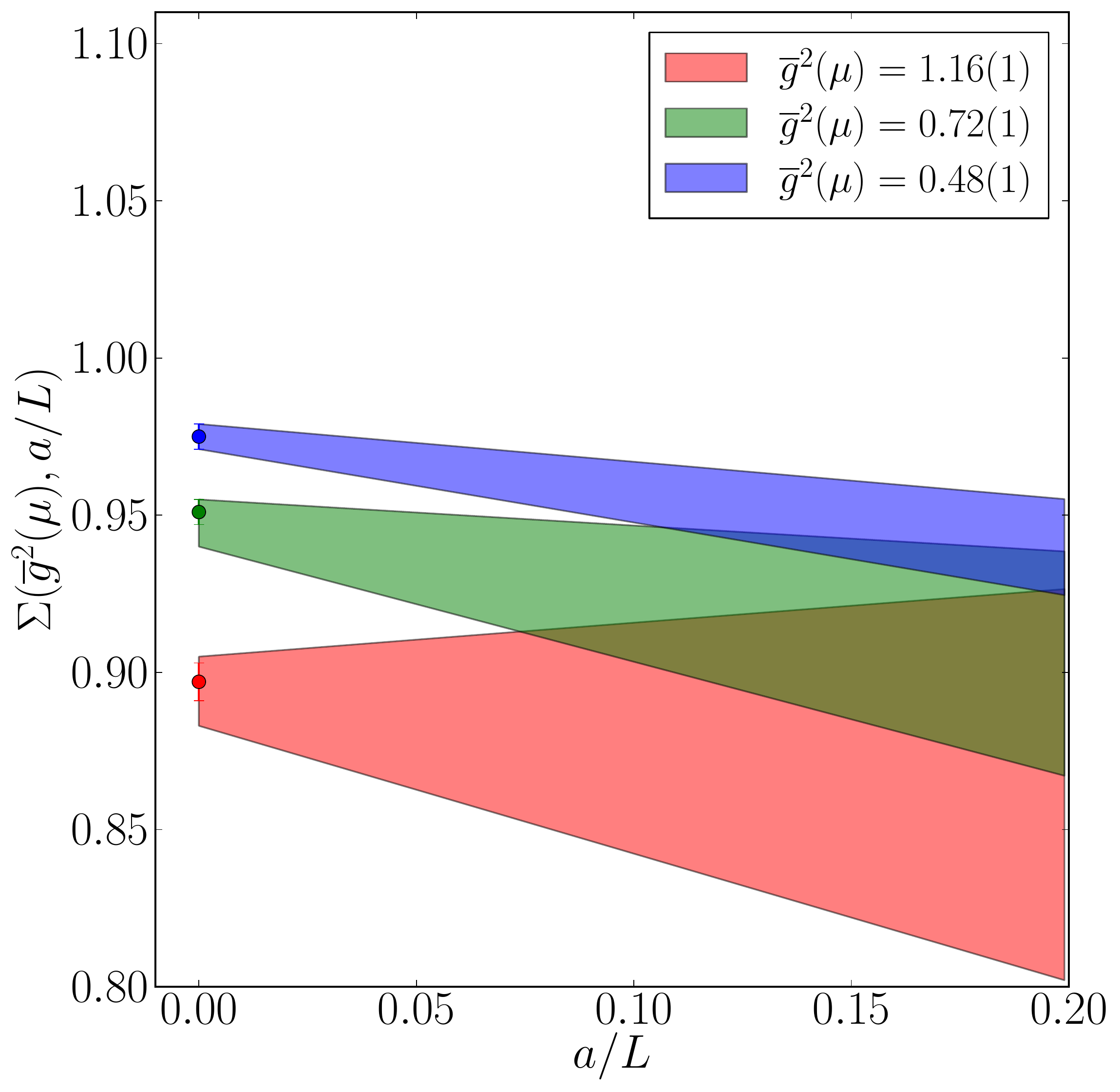}
\end{subfigure}
   \caption{Continuum extrapolation of the step-scaling function of the pseudo 
scalar 
density at several values of the renormalized gauge coupling with $c=0.3$. 
(Left) We plot the numerical data, with statistical uncertainties only, and 
include the continuum values as circular points. The lines are fits linear in 
$a/L$. (Right) Estimated 
systematic uncertainties. Estimates include both uncertainties from the 
continuum extrapolation and the critical mass mis-tuning, which are 
approximately 5-10\% for the lowest 
scale (in red), decreasing to less than 1\% for the highest scale (in blue). }
  \label{fig:sigma}
\end{figure}

\vspace*{-0.2\baselineskip}We plot the discrete 
step-scaling function, with the continuum limit at $a/L=0$, in the left-hand 
plot of Figure \ref{fig:sigma}. We include only systematic 
uncertainties in the plotted errorbars and in the fit. The red data 
points correspond to the lowest physical scale, the green to an intermediate 
scale and the blue to the highest physical scale. On the right-hand side of 
Figure \ref{fig:sigma} we illustrate our estimate for the systematic 
uncertainty from the continuum extrapolation and the critical mass 
mis-tuning. 

The spread of data points at each scale 
demonstrates that at higher scales the continuum extrapolation is 
better controlled. This is largely caused by the 
improved tuning of the critical mass at higher values of the bare coupling. The 
red data points correspond to bare coupling values between $\beta = 6.6$ and 
$\beta = 7.4$. As we discuss in the next section, these data suffer from 
systematic uncertainties due to mis-tuning of the critical mass that are 
approximately $5-10$\%. In contrast, the blue data have bare couplings in 
the range $\beta = 9.6$ to $\beta = 10.8$, with corresponding critical 
mass tuning uncertainties of less than 1\%.

\subsection{Tuning the critical mass}

We determined the critical mass for each value of the bare coupling constant 
using the two-loop cactus improved critical mass of \cite{Panagopoulos:2001fn}. 
We estimate that the resulting value of the critical mass is 
mis-tuned by approximately
18\% at $\beta = 6.0$, 15\% at $\beta = 6.1$ and 12\% at $\beta = 
6.3$, by comparing the perturbative predictions to the nonperturbative results 
quoted in~\cite{Follana:2000mn}. The smallest $\beta$ value we use is 6.6, for 
which we estimate a critical mass 
mis-tuning of approximately 10\%. Ref.~\cite{Panagopoulos:2001fn} suggests that 
this mis-tuning is reduced to $\ll$ 1\% for the highest values of $\beta$ in 
this work, $11 < \beta < 12$.

To estimate the resulting error in the renormalization parameter we determined 
$Z_p$ at $\beta =  6.1$ and $6.3$, using both the perturbative and 
nonperturbative values of the critical mass. We find that the resulting 
uncertainty is approximately 18\% and 13\% for $\beta =  6.1$ and 
$6.3$ respectively, Uncertainties for the $\beta = 6.0$ ensemble were too large 
to use for this analysis.

Nonperturbatively tuned values of the critical mass are not available at larger 
values of the bare coupling. We therefore estimated the uncertainties in the 
renormalization factor by varying the critical mass by around 10\% at higher 
beta values (8.0 to 12.0). This resulted in variations of $Z_p$ of around 1\%. 
Since the mis-tuning of the 
critical mass is likely to be significantly less than 10\% at these values of 
the bare coupling, we conclude that the systematic effects of the critical mass 
tuning are much less than 
1\% for the high $\beta$ range. At $\beta = 6.6$, we estimate the systematic 
uncertainty to be around 10\%.

\subsection{Other sources of uncertainty}

The systematic uncertainties due to the critical mass tuning dominate the 
uncertainties in $Z_p$ and consequently in $\sigma(\overline{g}^2(\mu))$, 
particularly at low scales. The 
continuum extrapolation, which is linear in the lattice spacing, because we use unimproved Wilson fermions, is the next 
most significant source of systematic error and the corresponding uncertainties 
are approximately 1-2\%. Statistical errors in $Z_p(g_0,\mu,a/L)$ were 
calculated
from jackknife estimates to include correlations between the correlation 
functions of the pseudoscalar and vector currents and were $\sim 0.5$\%. 
We determined that the bias in these uncertainties from the jackknife 
procedure was negligible. We estimated that uncertainties due to 
mis-tuning of the renormalized coupling were $\sim 0.5$\%.

\section{Conclusions}

We have introduced a new finite volume renormalization scheme. We define this 
scheme through two point correlation 
functions of smeared and local operators, using the Gradient Flow to smear both 
fermion and gauge fields. We remove any dependence on the wave function 
renormalization for smeared fermions by considering an appropriate ratio of 
correlation functions. Thus the renormalization parameter of this ratio
includes only the renormalization factor of the local operator. 
Our scheme therefore provides a simple way to compute the renormalization 
parameters of lattice operators. 

We adopt a normalization condition that restricts the path integral to constant 
gauge fields. We compute these constant gauge field correlation functions 
numerically. This condition simplifies the matching to the $\overline{MS}$ 
scheme. Furthermore, with our
scheme one can nonperturbatively determine the scale dependence of the local
operator using a step-scaling 
procedure. 

Here we have demonstrated the basic features of our scheme using Wilson 
fermions in the quenched approximation and concentrating on the pseudo-scalar 
quark bilinear operator. In the future we will pursue unquenched, $O(a)$ 
improved calculations. 

The ultimate goal of this work is to calculate 
nonperturbatively the renormalization constants, mixing coefficients and 
scale dependence of twist-2 matrix elements relevant to hadron structure 
calculations. Extracting phenomenologically relevant results requires matching 
to the $\overline{MS}$ scheme and work to carry out these matching calculations 
is underway. 

\bibliography{flow}{}

\bibliographystyle{unsrt}

\end{document}